\documentclass[12pt]{article}

\usepackage{amsmath}
\usepackage[psamsfonts]{amssymb}
\usepackage{amsthm}
\usepackage{cite}
\usepackage{indentfirst}
\usepackage{graphicx}

\newcommand{\N}{\mathcal{N}}
\newcommand{\Tr}{\mathrm{Tr\ }}
\newcommand{\LL}{\mathcal{L}}
\newcommand{\pert}{\mathrm{pert}}
\newcommand{\Z}{\mathbb{Z}}

\newcommand{\half}{\frac{1}{2}}
\newcommand{\abs}[1]{\lvert#1\rvert}
\newcommand{\Trs}{\mathrm{Tr}_{SU(N)}}
\newcommand{\Trss}{\mathrm{Tr}_{SU(2N)}}
\newcommand{\W}{\mathcal{W}}

\newcommand{\Dirac}{\!\!\not\!\!D}

\begin{document}

\bibliographystyle{utphys}

\begin{titlepage}

\begin{center}
\hfill HUTP-02/A059\\
\hfill ITFA-02-52\\
\hfill hep-th/0211194

\vskip 1.5 cm
{\large \bf Large $N$ Strong Coupling Dynamics in \\
         Non-Supersymmetric Orbifold Field Theories}
\vskip 1 cm 
{Robbert Dijkgraaf$^1$, Andrew Neitzke$^2$ and Cumrun Vafa$^2$}\\
\vskip 0.5cm
{\sl $^1$Institute for Theoretical Physics \& Korteweg-de Vries Institute for Mathematics,\\
University of Amsterdam, 1018 XE Amsterdam, The Netherlands\\}
\vskip 0.5cm
{\sl $^2$Jefferson Physical Laboratory,
Harvard University,\\
Cambridge, MA 02138, USA}
\end{center}

\vskip 0.5 cm
\begin{abstract}
We give a recipe relating holomorphic quantities in supersymmetric field theory
to their descendants in non-supersymmetric $\Z_2$ orbifold field theories.
This recipe, consistent with a recent proposal of Strassler, 
gives exact results for bifermion condensates, domain wall
tensions and gauge coupling constants in the planar limit of the orbifold theories.
\end{abstract}

\end{titlepage}

\setcounter{page}{1}
\pagestyle{plain}

\section{Introduction}

Over the last decade there has been enormous progress in our understanding of
supersymmetric field theories, aided by the discovery of non-perturbative 
dualities both in field theory and in string theory.  In general there has
been much less progress in the non-supersymmetric case.  However, there was
one notable exception where the dualities actually gave information about
a non-supersymmetric theory; namely, motivated from AdS/CFT, it was
observed in \cite{Kachru:1998ys, Lawrence:1998ja} that one should expect certain non-supersymmetric 
orbifolds of $\N=4$ super Yang-Mills to be conformal at least in the large $N$ limit.
Subsequently this statement was proven, first using string theory in \cite{Bershadsky:1998mb}
and more generally in field theory in \cite{Bershadsky:1998cb}.  The proof
consists in showing that the leading
diagrams in the large $N$ expansion satisfy a kind of ``inheritance principle,'' meaning they
are equal in parent and daughter theories (up to a rescaling of
the coupling constant.)  This proof in fact applies even
when one starts with a non-conformal theory.

Recently it was conjectured by Strassler \cite{Strassler:2001fs} that the inheritance principle
might hold even nonperturbatively in the 't Hooft coupling.
As he pointed out, if this conjecture is true, it is a powerful tool for obtaining 
nonperturbative information, exact in the large $N$ limit,
in some nonsupersymmetric theories --- namely those which can be obtained 
as orbifolds of $\N=1$ super Yang-Mills, which has been well understood nonperturbatively for a long time.  But
the conjecture is difficult to check; in \cite{Strassler:2001fs} it was suggested that the best avenue for testing
it would be lattice simulations.

On the other hand, it was recently shown \cite{Dijkgraaf:2002dh, Dijkgraaf:2002xd} that, in supersymmetric gauge theories 
with a large $N$
expansion, all holomorphic quantities --- including ones which receive nonperturbative corrections ---
can be calculated by minimizing a potential which
is calculated solely perturbatively and indeed receives contributions only from 
planar diagrams.  These planar diagrams are not precisely of the type 
considered in \cite{Bershadsky:1998mb, Bershadsky:1998cb}
but we will show that they nevertheless satisfy an inheritance principle for a similar reason
as long as we consider a specific class of $\Z_2$ orbifolds.
Using the techniques of \cite{Dijkgraaf:2002dh} one can then obtain 
exact results in the large $N$ limit of any nonsupersymmetric gauge theory obtained as a orbifold of a supersymmetric one;
more specifically, one should be able to compute any quantity in the orbifold theory which is descended from a 
holomorphic quantity in the parent.  In particular, we can determine the effect of additional
adjoint matter in the parent theory on the formation of condensates and $U(1)$ couplings
in the daughter.  This paper is a short description of how this works.

The organization of this paper is as follows.  In Section \ref{orbifold-theories} we review the notion
of orbifold field theory.  In Section \ref{result} we state our main result.  In Section \ref{diagrammatics} we give some diagrammatic 
motivation for the conjecture.  Finally in Section \ref{examples} we sketch a few examples of theories for which we can
make predictions using this technology.

\section{Orbifold theories} \label{orbifold-theories}

The field theories we will consider are obtained by a well-studied truncation procedure
motivated from string theory \cite{Douglas:1996sw}.  Namely, begin with some four-dimensional ``parent''
$U(N)$ gauge theory possessing a global symmetry $R$.  Then consider a finite subgroup $G$
of $U(N) \times R$, subject to the extra condition that the $U(N)$ part of any nontrivial element in $G$
must be trace-free in the fundamental representation of $U(N)$.  The truncated ``daughter'' theory is
then defined simply by setting to zero
all of the fundamental fields which are not $G$-invariant.
When the parent theory has a string theory realization on a stack of 
D-branes, this procedure is equivalent to taking an orbifold of the string theory; the
trace-free condition corresponds to the requirement that none of the branes are stuck
at fixed points.

It is known that the planar diagrams of the daughter theory are numerically equal to the planar diagrams
of the parent theory (up to a rescaling of the gauge coupling); 
this was shown first using string theory to organize the perturbation series \cite{Bershadsky:1998mb}
and later this proof was rewritten strictly within field theory \cite{Bershadsky:1998cb}.  So the two theories are
the same at large $N$, at least perturbatively in $g^2 N$.

\begin{figure}[t]
\centering
\includegraphics{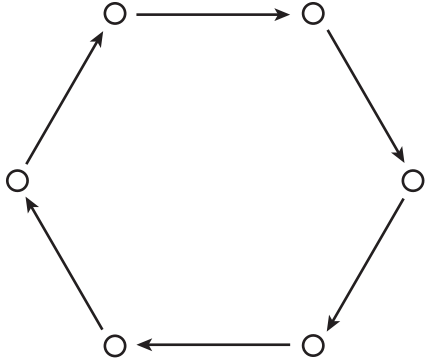}
\caption{The matter content of $\N=1$ SYM after a $\Z_p$ truncation (in the case $p=6$.)}
\label{circle-quiver}
\end{figure}

In this paper we will focus on examples 
where the parent theory is $\N=1$ supersymmetric
and the daughter is non-supersymmetric.  The simplest such example, also discussed in
some detail in \cite{Strassler:2001fs}, is the case where the parent is $\N=1$, $U(pN)$ super Yang-Mills and
$G = \Z_p$, embedded into the $U(1)$ R-symmetry and simultaneously 
acting by $N$ copies of the regular (cyclic permutation)
representation of $\Z_p$ on the fundamental representation of $U(pN)$.  In this case the daughter
theory has gauge group $U(N)^p$, and 
for each $i$ between $1$ and $p$ it has a massless Weyl fermion transforming in the fundamental of
$U(N)_i$ and the antifundamental of $U(N)_{i+1}$ (with the convention $p+1 = 1$).
This matter content is summarized by the quiver diagram in Figure \ref{circle-quiver}.  
(Throughout this paper all quiver diagrams
are $\N=0$ quivers; nodes represent gauge groups and lines with arrows represent bifundamental Weyl fermions
or scalars.)

We will consider only the case $p = 2$.  Pragmatically this restriction is motivated by our belief that
bifermion condensates will give good variables with which to describe the infrared dynamics; 
such a condensate can be gauge invariant only if $p = 2$.  But actually
there is another good reason to restrict attention to $p=2$, namely, this is the only case in which
the whole group $\Z_p$ preserves the vacuum of the parent theory (recall that in $\N=1$, $U(pN)$ SYM
the R-symmetry is broken $U(1) \to \Z_{2pN} \to \Z_2$, first by an anomaly and then spontaneously by
the choice of one of $pN$ vacua.)  One would expect that
the only effect of the truncation by elements which do not preserve the vacuum is to identify the 
different vacua, not to change the physics in a particular vacuum.  So if $p$ is odd then none of the
group elements affect the physics, and the theory should still be equivalent 
to $\N=1$ SYM; if $p$ is even only the element of order $2$ affects the physics, so the theory
should be equivalent to the $\Z_2$ quotient.  Indeed, at least if one is allowed 
to adjust the gauge couplings away from the orbifold point, this is known to be the case; confinement and
chiral symmetry breaking can reduce $p$ to $p-2$ repeatedly until one is left with either $p=1$ or $p=2$ in the infrared
depending on whether one started with $p$ odd or even respectively 
\cite{Strassler:2001fs, Georgi:1986hf}.

\begin{figure}[t]
\centering
\includegraphics{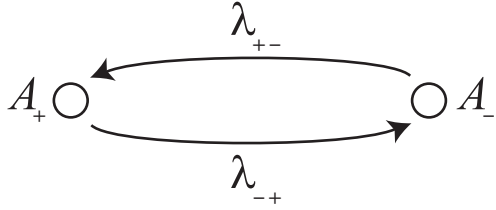}
\caption{The matter content of the $\N=1$ vector multiplet after the $\Z_2$ truncation by \eqref{twist}.}
\label{simple-quiver}
\end{figure}

So we will consider a $\Z_2$ truncation of an $\N=1$, $U(2N)$ gauge theory,
with the nontrivial group element acting as
\begin{equation} \label{twist}
(-1)^F \cdot \begin{pmatrix} 0_{N \times N} & 1_{N \times N} \\ 1_{N \times N} & 0_{N \times N} \end{pmatrix}
\end{equation}
where the matrix, acting on the gauge indices, is written in $N \times N$ blocks.  The matter
content of the theory will include the truncation of the $\N=1$ 
vector multiplet; this yields two $U(N)$ gauge fields $A_+$, $A_-$, with equal holomorphic gauge couplings,
and two bifundamental fermions
$\lambda_{+-}$, $\lambda_{-+}$ as shown in Figure \ref{simple-quiver}.

If the parent theory has additional
matter fields they will give rise to additional matter in the daughter theory; this is the case in
which the matrix model techniques of \cite{Dijkgraaf:2002dh} will be most useful.

\section{Results} \label{result}

We are now ready to state our main results.  We consider the $\Z_2$ quotient \eqref{twist} of 
$\N=1$, $U(2N)$ super Yang-Mills with additional
adjoint matter and a superpotential with isolated classical solutions.
To predict the nonperturbative effects of the daughter theory at large $N$ we will take our cue from the
result of \cite{Dijkgraaf:2002dh}, which showed that the vacuum structure and holomorphic data of the parent
can be calculated purely perturbatively.  Specifically, \cite{Dijkgraaf:2002dh} gave a general prescription
for the infrared dynamics of the chiral superfield $S = \frac{1}{32 \pi^2} \Trss \W\W$:  namely,
the effective superpotential is given by
\begin{equation} \label{parent-W}
W(S) = 2NS \log (S / \Lambda_0^3) - 2 \pi i \tau S + W_{\pert}(S),
\end{equation}
where $W_{\pert}(S)$ is computed from the planar diagrams of a matrix model determined by the tree level superpotential of the
parent theory, and
\begin{equation}
\tau = \frac{\theta}{2\pi} + \frac{4 \pi i}{g^2}
\end{equation}
is the holomorphic bare coupling.
In component fields the superfield $S = s + \theta \chi + \theta^2 G$ contains the gaugino condensate $s$
as well as the auxiliary field $G = \frac{1}{64 \pi^2} \Tr (F \wedge *F - i F \wedge F)$, which is essential for the
reproduction of the chiral anomaly.  Integrating over the auxiliary field $G$ then gives the scalar potential
$\abs{W'(s)}^2$.

We propose that the daughter theory at large $N$ 
admits a similar infrared description.
Namely, the daughter has a possible bifermion condensate
$s_d = \frac{1}{16 \pi^2} \Trs \lambda_{+-} \lambda_{-+}$, obtained by truncating
$s = \frac{1}{32 \pi^2} \Trss \lambda \lambda$ to its invariant constituents,
and we can also write $G_d$ for the truncation of $G$.  Then by strictly perturbative
calculations reviewed in Section \ref{diagrammatics} one can see that the effective
action for $s_d$ and $G_d$ contains terms which are naturally written 
$G_d W'_d(s_d)$ for a function $W_d(S_d)$.

Furthermore, as we will show, $W_d$ is obtained directly from $W$ by matching up quantities
in the parent and daughter theory.  First, we should identify
\begin{equation}
S_d = 2S.
\end{equation}
From perturbative calculations \cite{Bershadsky:1998mb, Bershadsky:1998cb} we
also know that the free energies in parent and daughter can be equal only if
one identifies $g_d^2 = 2 g^2$, and we propose that the correct nonperturbative extension
of this is
\begin{equation}
\tau_d = \tau / 2.
\end{equation}
Finally, there is an inheritance principle
\begin{equation}
W_d = W.
\end{equation}
Substituting these in \eqref{parent-W} one finds
\begin{equation} \label{daughter-W}
W_d(S_d) = NS_d \log (S_d / 2 \Lambda_0^3) - 2 \pi i \tau_d S_d + W_{\pert}(S_d/2).
\end{equation}

Of course, this $W_d(S_d)$ does not admit a simple interpretation as a superpotential.
Nevertheless we propose that it continues to play the usual roles played by the
superpotential ---
in particular, the vacua of the daughter theory at large $N$
are the critical points of $W_d(S_d)$, and domain wall tensions are
given by differences of $W_d$.  (Essentially this is equivalent to
saying that, at least at large $N$, the ``D-terms'' are sufficiently
benign that $G_d$ can be treated as an auxiliary
field and integrated out.)

We can also give a formula for the diagonal $U(1)$ couplings of the daughter theory,
essentially $\tau_d^{U(1)} = \tau^{U(1)} / 2$, which will be discussed in more detail below.

\section{Diagrammatics} \label{diagrammatics}

In this section we give the diagrammatic derivations of $W_d(S_d)$ and $\tau^{U(1)}(S_d)$.
In the spirit of \cite{Bershadsky:1998mb, Dijkgraaf:2002dh} we will exploit the string theory 
representation of field theory diagrams as an organizing
tool in our arguments; however, all the arguments can be rephrased purely in terms of 
field theory a la \cite{Bershadsky:1998cb, Dijkgraaf:2002xd}.

\subsection{Diagrammatics of $W(S)$}

The diagrams which contribute to the computation of $W(S)$ in 
the parent $\N=1$ theory are
slightly different from the planar diagrams considered
in \cite{Bershadsky:1998mb, Bershadsky:1998cb}.  Namely, those diagrams only had open string insertions 
along at most one boundary.  They are the diagrams which have the leading
$N$ dependence, and for such diagrams it is easy to see 
that, after the $\Z_2$ orbifolding, twisted sectors cannot contribute to
the path integral; namely, any twisted sector
will have a twist along at least one boundary without insertions,
but then the Chan-Paton trace along that boundary gives zero.

\begin{figure}[t]
\centering
\includegraphics{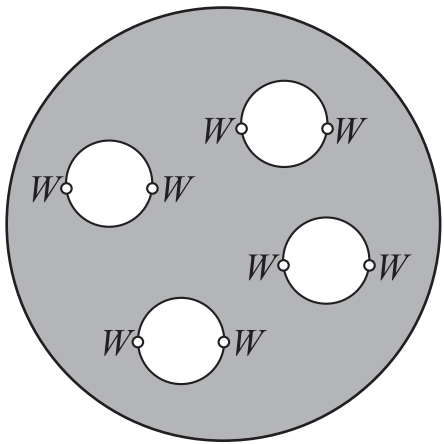}
\caption{A diagram contributing to $W_{\pert}(S)$.}
\label{w-diagram}
\end{figure}

\begin{figure}[t]
\centering
\includegraphics{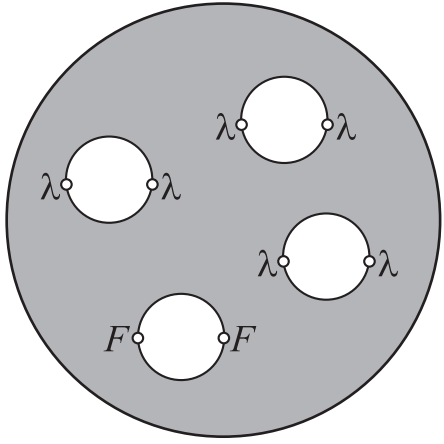}
\caption{A diagram contributing to $W_{\pert}(S)$, in superfield components.}
\label{w-diagram-components}
\end{figure}

On the other hand, to compute $W_{\pert}(S)$, according to \cite{Bershadsky:1994cx} one must insert two 
$\W$ background fields along each of $h-1$ boundaries as pictured in Figure \ref{w-diagram}.
We want to compare this diagram with its counterpart in the daughter theory.  In the daughter
theory the background fields we can turn on are more restricted since they must be invariant
under $\Z_2$.  Since the twist acts differently
on the different components of $S$, we should look at the diagram in terms of component fields; then it has
two $\lambda$ insertions on $h-2$ boundaries, two $F$ insertions on one boundary, and no insertions
on one boundary, as in Figure \ref{w-diagram-components}.
If there is a twist on the boundary with no insertions we get zero, because $\Tr(\gamma) = 0$ where
\begin{equation}
\gamma = \begin{pmatrix} 0_{N \times N} & 1_{N \times N} \\ 1_{N \times N} & 0_{N \times N} \end{pmatrix}.
\end{equation}
What if there is a twist on a boundary with $\lambda$ insertions?
Since $\lambda$ is fermionic and the twist includes a factor $(-1)^F$, for the background field $\lambda$ 
to be invariant its Chan-Paton part (which we also write $\lambda$) must anticommute with $\gamma$.  Then the relevant
trace in the twisted sector is 
\begin{equation}
\Tr(\lambda \lambda \gamma) = -\Tr(\lambda \gamma \lambda) = -\Tr(\lambda \lambda \gamma)
\end{equation}
where we used the cyclicity of the trace.  So any sector with a twist either on the empty boundary or on one
of the boundaries carrying $\lambda$ insertions gives zero.  The full value of the diagram is obtained by summing
over all allowed combinations of twists; there are $2^{h-1}$ of these, obtained by distributing the twists 
arbitrarily over the
$h$ boundaries subject to the constraint that the total number of twists is even.  From the above we then
see that only the completely untwisted sector can contribute.
Furthermore there is an overall rescaling factor $1 / 2^{h-1}$ for the sum over $2^{h-1}$ twisted sectors.
This factor we can interpret as giving $1/2$ for every boundary with an $S$ insertion; hence in the daughter
theory we have to set $S = S_d / 2$, at least in $W_\pert$.

Actually, we should make this replacement $S = S_d / 2$
everywhere in $W$.
One way to see this is to observe that after this replacement $2NS \log S$ becomes $NS_d \log S_d$ giving the
correct chiral anomaly in the daughter theory; another way is to note that the division between the 
logarithmic term and $W_\pert$ is unnatural from the viewpoint of the matrix model, in which the logarithm
arises from the path integral measure.

\subsection{Diagrammatics of $U(1)$ couplings}

If the gauge symmetry of the parent $U(2N)$ theory is unbroken and we have
only adjoint fields, then in the parent the only
$U(1)$ factor is the overall $U(1)$ which is decoupled.  In the daughter
$U(N) \times U(N)$ theory there is a $U(1) \times U(1)$; one combination of these is the 
trivial decoupled one, and as we will see below, the inheritance principle
does not determine
the coupling constant for the other combination.  So to obtain an interesting result
for the $U(1)$ couplings we must consider a more general situation.  Indeed, 
for a general choice of $\N=1$ parent theory we can consider a vacuum with the
symmetry breaking pattern $U(2N) \to U(2N_1) \times \cdots \times U(2N_n)$.  Choosing
our $\Z_2$ quotient to respect this breaking, we will find daughter vacua with gauge
group $(U(N_1) \times U(N_1)) \times \cdots \times (U(N_n) \times U(N_n))$.  
In the infrared we will have confinement for the nonabelian factors, leaving behind
$U(1)^{2n}$.

\begin{figure}[t]
\centering
\includegraphics{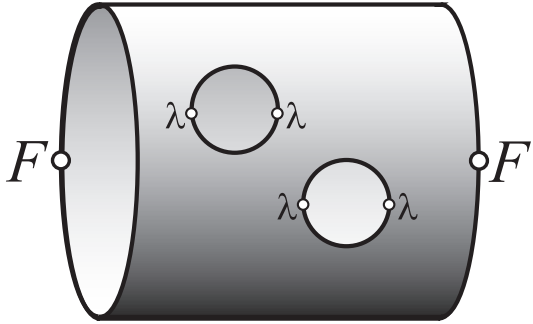}
\caption{A diagram contributing to $\tau^{U(1)}(S)$.}
\label{tau-diagram}
\end{figure}

The matrix of effective $U(1)$ couplings in the
parent theory is obtained from diagrams
which have two boundaries with one $F$ each, and $h-2$ boundaries with two $\lambda$ each, as shown in Figure
\ref{tau-diagram}.
More explicitly, the rule is that for any background fields $F_i$, the sum of these diagrams with the $F_i$ inserted
gives $- 2 \pi i \tau_{ij}^{U(1)}(s) (\Tr F_i) (\Tr F_j)$.

As in the previous subsection, in the daughter theory the sectors with twists on the
boundaries with $\lambda$ insertions give zero.
But depending on which components of $F$ we turn on, there
may or may not be contributions from the sector with a twist on both ends of the cylinder
(in other words, the $F$ insertions can create both twisted and untwisted sector
states in the closed string channel, as already observed
in \cite{Bershadsky:1998mb}.)  Let us organize the $2n$ $U(1)$ field strengths of the daughter theory
into $F_{iu} = F_{i+} + F_{i-}$, $F_{it} = F_{i+} - F_{i-}$.  The notation is explained by the fact that 
on boundaries with an $F_{iu}$ inserted only the untwisted sector contributes, and on boundaries with
an $F_{it}$ inserted only the twisted sector contributes.  In the basis 
$\{ F_{iu}, F_{it} \}$ the $U(1)$ couplings of the daughter
theory are therefore 
\begin{equation}
\begin{pmatrix} \half \tau^{U(1)}_{ij} & 0 \\ 0 & \half \tau'^{U(1)}_{ij} \end{pmatrix}
\end{equation}
where $\tau^{U(1)}_{ij}$ are the $U(1)$ couplings in the parent theory, given by the untwisted sector diagrams
and explicitly computable as in \cite{Dijkgraaf:2002dh},
while $\tau'^{U(1)}_{ij}$ come from the corresponding twisted sector diagrams.  Hence it is only the diagonal
$U(1) \subset U(N_i) \times U(N_i)$ for which the couplings are directly calculable from the inheritance
principle.  (One might have thought that the condition for avoiding contributions from twisted sectors
would simply be $\Tr (\gamma F) = 0$, which would only exclude one of the $2n$ $U(1)$ couplings
rather than $n$ of them; this is wrong because in the vacuum with broken gauge symmetry the dependence on
the Chan-Paton index is not just an overall factor $\Tr (\gamma F)$.)

\section{Some examples} \label{examples}

The simplest case is the case where the parent is pure $\N=1$ super Yang-Mills.
In this case the parent simply has the Veneziano-Yankielowicz superpotential \cite{Veneziano:1982ah} dictated by the one-loop axial anomaly,
\begin{equation}
W(S) = 2NS \log (S / \Lambda_0^3) - 2 \pi i \tau S,
\end{equation}
which on extremization gives the standard $2N$ vacua determined by
\begin{equation}
( S / \Lambda_0^3 )^{2N} = e^{2 \pi i \tau}.
\end{equation}
The daughter theory under the $\Z_2$ action \eqref{twist} 
is as pictured in Figure \ref{simple-quiver}, with all couplings equal ($g_+ = g_- = g_d$, $\theta_+ = \theta_- = 
\theta_d$.)
So in this case we have simply
\begin{equation}
W_d(S_d) = NS_d \log (S_d / 2 \Lambda_0^3) - 2 \pi i \tau_d S_d
\end{equation}
which on extremization gives $N$ vacua determined by
\begin{equation}
( S / 2 \Lambda_0^3 )^{N} = e^{2 \pi i \tau_d}
\end{equation}
or more explicitly,
\begin{equation}
S = 2 \Lambda_0^3 e^{-2 \pi i \tau_d / N} e^{2 \pi i k / N} = 2 \Lambda_0^3 e^{- 8 \pi^2 / {g_d^2}N} e^{i \theta_d / N} e^{2 \pi i k / N}.
\end{equation}
This result is consistent with the chiral anomaly of the daughter theory; under a chiral rotation
by $e^{i \alpha}$, $S$ must rotate by $e^{2 i \alpha}$, while $\theta$ shifts by $2N \alpha$.  
It is also consistent with the large $N$ RG flow, since at one loop the dynamical scale is 
$\Lambda = \Lambda_0 e^{- 8 \pi^2 / 3{g^2}N}$ and this one-loop
result is exact at large $N$, at least perturbatively in $g^2N$ \cite{Bershadsky:1998cb}.  Put another
way:  independent of the way we constructed the daughter theory, 
its RG flow and chiral anomaly together imply that the condensate depends holomorphically 
on $\tau_d$ at large $N$.  This is a special feature of orbifolds of $\N=1$, not shared by generic $\N = 0$ gauge theories;
we believe it is only because of this holomorphicity that we have any chance of getting exact results.

Note that because $\theta_d = \theta / 2$ there is an ambiguity of $\pi$, rather than the usual $2\pi$, 
in our definition of $\theta_d$.  This ambiguity can
shift $S$ by a factor $e^{i \pi / N}$ and is related to the fact that the parent theory had $2N$ vacua while the 
daughter only has $N$.

One can obtain more complicated examples by adding chiral superfields $\Phi$ to the parent.
With no superpotential this would give $\N = 2$ super Yang-Mills and a continuous moduli space; 
we add a superpotential $\Tr W(\Phi)$ to break back down to $\N = 1$ and lift the moduli.
The simplest possibility is to add just one adjoint superfield.  For any
choice of $W(\Phi)$ these models have been exactly solved in \cite{Cachazo:2001jy}, so the exact value of $W_{\pert}(S)$
is known.

\begin{figure}[t]
\centering
\includegraphics{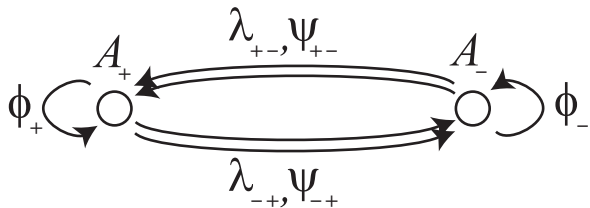}
\caption{The matter content of the daughter of $\N=1$ SYM with one adjoint chiral superfield after the $\Z_2$ truncation by \eqref{twist}.}
\label{adjoint-quiver}
\end{figure}

Then the matter content of the daughter theory
looks like Figure \ref{adjoint-quiver},
while the action is obtained by truncation of the parent Lagrangian, giving
\begin{align} \label{daughter-L}
\LL =\ \Tr\ & \Big( \half (\abs{D^\mu \phi_+}^2 + \abs{D^\mu \phi_-}^2 + \frac{1}{4g_d^2} (F_+^2 + F_-^2) \nonumber \\
&+ \bar{\psi}_{+-} \Dirac \psi_{+-} + \bar{\psi}_{-+} \Dirac \psi_{-+} + \Big\lvert \frac{\partial W(\phi_+)}{\partial \phi_+} \Big\rvert^2 + \Big\lvert \frac{\partial W(\phi_-)}{\partial \phi_-} \Big\rvert^2 \\
&+ \Big( \phi_+(\lambda_{+-}\psi_{-+} - \psi_{+-}\lambda_{-+}) + \phi_-(\lambda_{-+}\psi_{+-} - \psi_{-+}\lambda_{+-}) \nonumber \\
&- \half \frac{\partial^2 W(\phi_+)}{\partial \phi_+^2} \psi_{+-} \psi_{-+} - \half \frac{\partial^2 W(\phi_-)}{\partial \phi_-^2} \psi_{-+} \psi_{+-} + c.c. \Big) \Big). \nonumber
\end{align}
A simple example is
\begin{equation}
W(\Phi) = \frac{m}{2} \Phi^2 + \frac{g}{3} \Phi^3.
\end{equation}
First note that in case $m \to \infty$, $\Phi$ is simply decoupled and we recover pure $\N=1$ SYM. 
Moreover, the result of \cite{Cachazo:2001jy} allows us to integrate out $\Phi$ and 
obtain the exact effective potential for $S$ even when $m$ is finite.  
Then the formula \eqref{daughter-W} for $W_d(S_d)$ determines the vacua and domain wall tensions 
of the daughter theory.

\begin{figure}[t]
\centering
\includegraphics{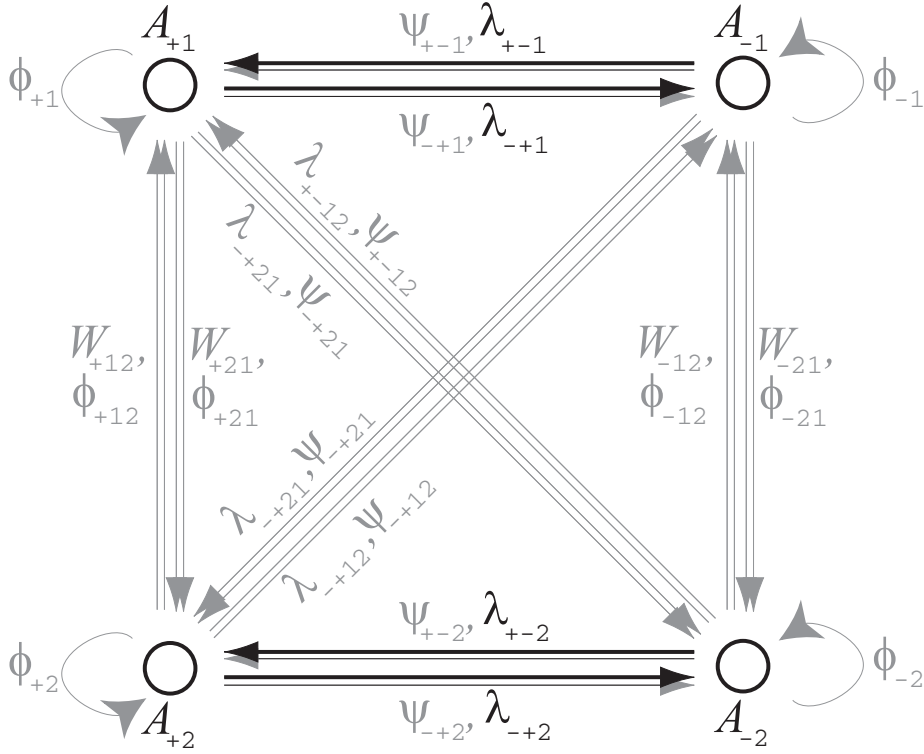}
\caption{The daughter of $\N=1$ SYM with one adjoint superfield and a superpotential $W(\Phi) = g(a^2 \Phi - \frac{1}{3} \Phi^3)$, 
after the adjoint gets a vev.  The gray lines represent massive fields.}
\label{double-quiver}
\end{figure}

We can also consider examples where $\Phi$ gets a vacuum expectation value.
The simplest such is obtained by taking
\begin{equation}
W(\Phi) = g(a^2 \Phi - \frac{1}{3} \Phi^3).
\end{equation}
In this case both $\phi_+$ and $\phi_-$
will get vacuum expectation values.  Since $W'(\pm a) = 0$, each of $\phi_\pm$ can separately distribute its
$N$ eigenvalues between $a$ and $-a$.  However, the only vacua for which we expect to be able to make a prediction
from the inheritance principle are the ones in which $\phi_+$ and $\phi_-$ distribute their eigenvalues equally, 
say each with $N_1$ eigenvalues equal to $a$ and
$N_2$ eigenvalues equal to $-a$ ($N_1 + N_2 = N$.)  These vacua descend from vacua with gauge symmetry
$U(2N_1) \times U(2N_2)$ in the parent theory,
where we have made the $\Z_2$ quotient in a way compatible with the gauge symmetry breaking.  
Substituting these vacuum expectation values
for $\phi_+$ and $\phi_-$ in \eqref{daughter-L} one obtains a theory with a complicated matter content shown
in Figure \ref{double-quiver}.
It describes two copies of the theory we considered in the first example (Figure \ref{simple-quiver}),
each with extra matter fields with masses of order $g$.  These two theories are then
coupled to one another by scalars, fermions and W bosons all with masses of order $a$.  So now the effective description
is in terms of two gluino condensates $S_{1d}$, $S_{2d}$, which are coupled to one another 
in $W_d(S_{1d}, S_{2d})$ (with couplings suppressed by powers of $a$.)  Specifically,
in the parent theory the superpotential is of the form
\begin{equation}
W(S_1, S_2) = \sum_i 2N_i S_i \log (S_i / \Lambda_0^3) - 2\pi i \tau S_i + W_{\pert}(S_1, S_2),
\end{equation}
so in the daughter we predict
\begin{equation}
W_d(S_{1d}, S_{2d}) = \sum_i \left(N_i S_{id} \log(S_{id} / 2 \Lambda_0^3) - 2\pi i \tau_d S_{id}\right) + W_{\pert}(S_{1d}/2, S_{2d}/2).
\end{equation}
We can also compute the couplings for the diagonals $U(1)_1 \subset U(N_1) \times U(N_1)$ 
and $U(1)_2 \subset U(N_2) \times U(N_2)$, which descend from the parent theory as
\begin{equation}
\tau_{ijd}^{U(1)} = \half \tau_{ij}^{U(1)}.
\end{equation}

\section*{Acknowledgements}

We thank Mina Aganagic, Nima Arkani-Hamed, Sergei Gukov and David Tong for useful discussions.
  
The research of R.D. is partially supported by FOM and the CMPA grant of the University of Amsterdam;
A.N. is supported by an NDSEG Graduate Fellowship;
C.V. is partially supported by NSF grants PHY-9802709 and DMS-0074329.

\providecommand{\OO}{} \renewcommand{\OO}{{\mathcal O}} \providecommand{\N}{}
  \renewcommand{\N}{{\mathcal N}} \providecommand{\PP}{}
  \renewcommand{\PP}{{\mathbb P}} \providecommand{\LL}{}
  \renewcommand{\LL}{{\mathcal L}} \providecommand{\Z}{}
  \renewcommand{\Z}{{\mathbb Z}} \providecommand{\g}{}
  \renewcommand{\g}{{\mathfrak g}}\def\cprime{$'$} \def\cprime{$'$}
\providecommand{\href}[2]{#2}\begingroup\raggedright\endgroup

\end{document}